\documentclass{jpp}
\usepackage{graphicx}
\usepackage{epstopdf, epsfig}
\usepackage[utf8]{inputenc}
\usepackage[T1]{fontenc}
\usepackage{amsmath}
\usepackage{algorithm}
\usepackage{tabularx}
\usepackage{algpseudocode}
\shorttitle{Image Classification With Cosine Similarity}
\shortauthor{Falato, Wolfe, Natan, Zhang, Marshall, Zhou, Bellan, and Wang}

\title{Plasma Image Classification Using Cosine Similarity Constrained CNN}
\author{Michael J. Falato\aff{1}, Bradley T. Wolfe\aff{1}, Tali M. Natan\aff{1}, Xinhua Zhang\aff{1}, Ryan S. Marshall\aff{2}, Yi Zhou\aff{2}, Paul M. Bellan\aff{2}
 \and Zhehui Wang\aff{1} \corresp{\email{zwang@lanl.gov}}}

\affiliation{\aff{1}Los Alamos National Laboratory, Los Alamos,
 NM 87545, USA 
 \aff{2}The California Institute of Technology, 1200 E California Blvd, Pasadena, CA 91125, USA}
\begin{document}
\maketitle
\begin{abstract}
Plasma jets are widely investigated both in the laboratory and in nature. Astrophysical objects such as black holes, active galactic nuclei, and young stellar objects commonly emit plasma jets in various forms. With the availability of data from plasma jet experiments resembling astrophysical plasma jets, classification of such data would potentially aid in investigating not only the underlying physics of the experiments but the study of astrophysical jets. In this work we use deep learning to process all of the laboratory plasma images from the Caltech Spheromak Experiment spanning two decades. We found that cosine similarity can aid in feature selection, classify images through comparison of feature vector direction, and be used as a loss function for the training of AlexNet for plasma image classification. We also develop a simple vector direction comparison algorithm for binary and multi-class classification. Using our algorithm we demonstrate 93\% accurate binary classification to distinguish unstable columns from stable columns and 92\% accurate five-way classification of a small, labeled data set which includes three classes corresponding to varying levels of kink instability.
\end{abstract}
\section{Introduction}
Big data and machine learning are monumental to the furthering of science. With many existing large image data sets \citep{imagenet, mnist}, machine learning and deep learning have become extremely effective at classifying such data \citep{dai2021coatnet, an2020ensemble}. Data sets for classification consist of different object types, or classes, and the true class labels for the images are known as the "ground truth". The task of classifying images given the ground truth of a training data set is known as supervised classification. In this study we perform classification of greyscale images from the Caltech Spheromak Experiment \citep{bellan}, where plasma jets and spheromaks are created and photographed in a vacuum chamber. Initial classification of data from this experiment is desirable in that it provides a starting point for further investigation into machine recognition of the underlying physics in the experiment from optical images, as well as extrapolation from laboratory plasma jets to our understanding of astrophysical plasma jets.

Convolutional neural networks (CNNs) are known to excel at supervised image classification \citep[see][]{cnn_rev}, where they achieve state of the art accuracy on benchmarks such as the Imagenet benchmark \citep{imagenet_bm}. Feature engineering is another common practice in machine learning and classification, where hand-defined, or engineered, features of images are often used in classification tasks to differentiate between images. The values of these features are organized into an array, which is used as a feature vector representing an image in the feature space. This study classifies both engineered feature vectors and output feature vectors from CNNs through the angle between feature vectors or similarly, the cosine of the angle between feature vectors known as the cosine similarity. Figure 1 displays a mock example of how this is done.

\begin{figure}
  \centering
  \includegraphics[scale = 0.8]{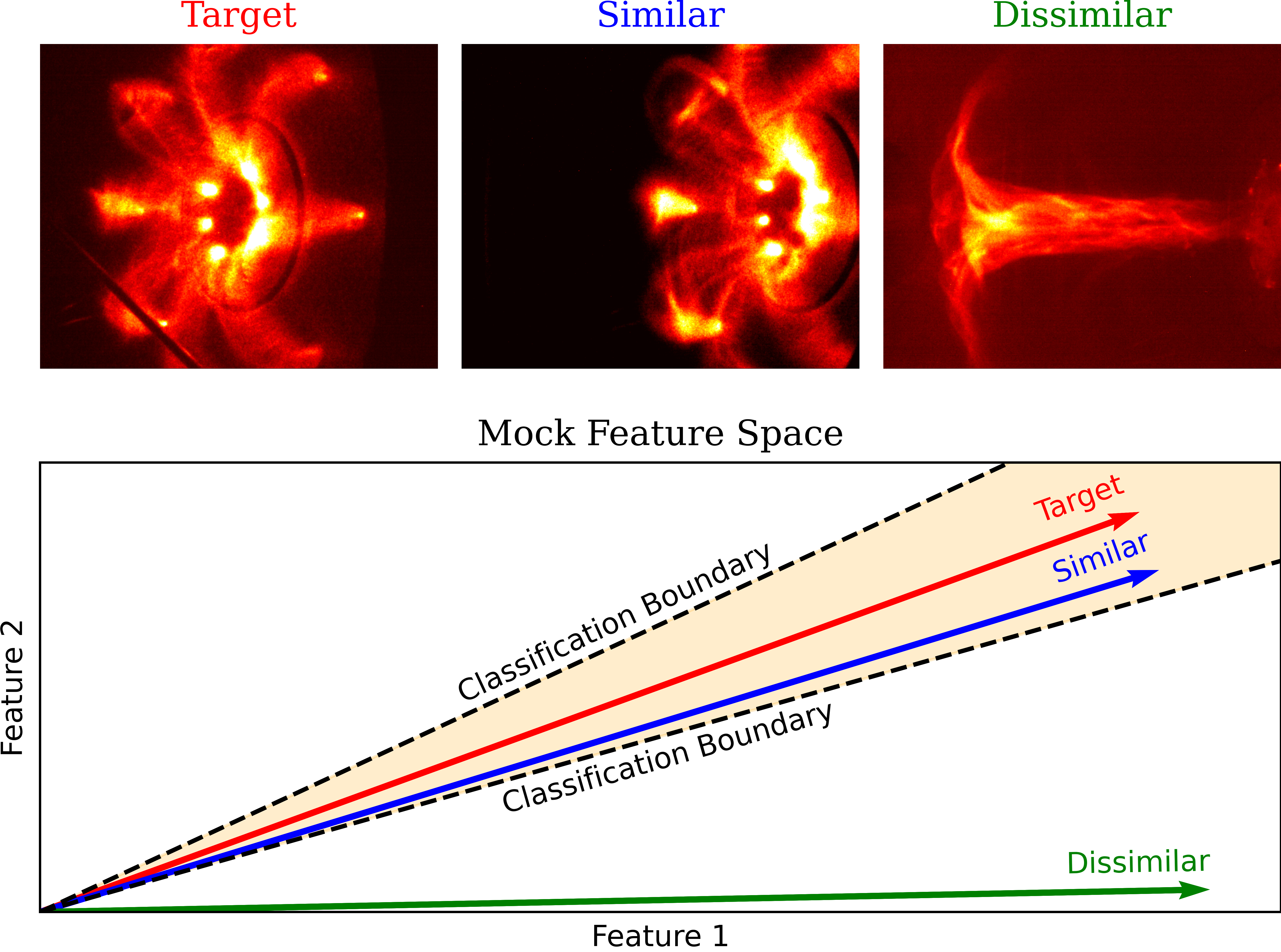}
  \caption{We conduct classification using the cosine similarity between feature vectors that are extracted from images. In order for classification to be conducted, the vectors must greatly align for images in the same class compared to images in different classes. Classification reduces to probing if a vector lies within a particular range in the binary classification case (shown above) or choosing the closest target region in the multi-class classification case. In this example, the similar image would be classified in the same class as the target image, while the dissimilar image would be classified out of the class.}
\label{fig:1}
\end{figure}

Cosine similarity has been used in face verification \citep{Nguyen2010CosineSM}, it can replace the dot product between output layers and weight vectors before activation layers in multi-layer networks \citep{luo2017cosine}, and it has been used to select robust feature subsets in genomic data analysis \citep{cs_gen}. Cosine similarity has also been used to adaptively refine clustering results in a pair-wise binary classification scheme \citep{dac}, and it is used in natural language processing to compare word embeddings \citep[see][]{young2018recent}. Cosine similarity has not yet been used to classify laboratory plasma images, and an algorithm that performs well using it to classify feature vectors is desirable given its simplicity. In this work we describe laboratory plasma image classification using cosine similarity. 

In addition, this study uses AlexNet \citep{alexnet}, one of the most well known CNNs, to achieve notably accurate transfer learning. Furthermore, we use cosine similarity of image feature vectors as a loss function for our instance of AlexNet, to select relevant features, and to classify images directly. If cosine similarity shows reasonable results in the ability to characterize similarity/difference between vectors in different classes based on specific features, then it should also be able to perform in a simple classification algorithm using vector direction comparison akin to the mock example from Figure 1. This work validates this idea by selecting features that distinguish the feature vector directions for images in a specific class from others. We train an instance of AlexNet to increase the cosine similarity of output feature vectors of images in the same class while attempting to orthogonalize feature vectors of images in different classes. We find that using a simple direction comparison algorithm with the output vectors from this instance of AlexNet achieves classification accuracies notably higher than traditional transfer learning with cross entropy. 

The remainder of this work is organized as: Methods and Results, where we give the details of the algorithms and image processing procedures while noting results obtained in using cosine similarity for feature selection and classification with AlexNet, Discussion, where the results and implications are assessed and future avenues are discussed, and Summary, where the key ideas from the paper are summarized.

\section{Methods and Results}
\subsection{Data Sets}
Our goal is to classify plasma jet images from the Caltech Spheromak Experiment \citep{scott}, where laboratory plasma jets are created and rich phenomena including kink instabilities are observed. Thus, our data sets consist of images sorted from this experiment. We create two data sets, one small data set (about 1000 images) with ground truth labels and one large data set (about 45000 images) with no ground truth labels.

For our small data set we sort the data into five classes, pertaining to five types of images that appear frequently in the experiment. Figure 2 shows a representative image from each class as an example, with a false colour map on each greyscale image. The classes are as follows: Haze, which corresponds to a general diffuse and spread-out plasma configuration, Spider, which corresponds to an initial eight plasma arcs tracing a background poloidal magnetic field occurring near the inception of a plasma jet, Column, a straight plasma jet showing little kink instability, Kink, an unstable column showing kink instability akin to a corkscrew shape, and Sphere, corresponding to anomalous spherical arrangements of pixels and plasma columns breaking off of the jet to form spheromaks. The number of images in each class for this small data set is 226, 111, 130, 295, and 390 for Haze, Spider, Column, Kink, and Sphere, respectively. We split each class in this set with a randomly distributed fractional split of 0.5, 0.25, and 0.25 and compile the corresponding splits of each class for training, validation, and test sets, respectively.

For our large data set, we separate images from the total data of about 300000 raw images using an instance of ResNet152 \citep{resnet} to distinguish between images based on quality.
We select 45000 high quality images for our large, unlabeled data set.
We set aside this large data set without labels for use in section 2.6 at the end of this work.
 For a thorough introduction to the experiment yielding these images and the theory involved, see \citep{bellan}.
\begin{figure}
  \centering
  \includegraphics[scale = 0.9]{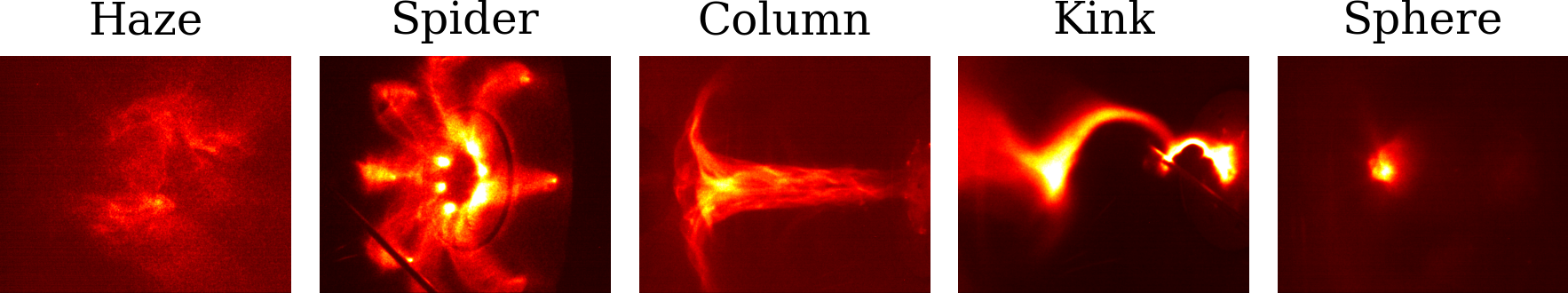}
  \caption{An example image from each of the five classes in the small data set. The classes are Haze, Spider, Column, Kink, and Sphere. Each class contains 226, 111, 130, 295, and 390 images respectively. Note that the images feature a false colour map.}
\label{fig:2}
\end{figure}

\subsection{Feature Selection Using Cosine Similarity}
To classify our small data set, we first turn to statistical and information theoretic features of images. We conduct feature extraction on our data set, inspired by the ability to view a greyscale image as a two dimensional intensity distribution. We transform each image into four statistical distributions, $P(x,y)$, $P(x)$, $P(y)$, and $P(v)$, for the collection of features from these distributions.  An example of these distributions obtained from an image can be seen in Figure 3, which displays an image, projections to the three distributions, $P(x,y)$, $P(x)$, and $P(y)$, and an example of the $P(v)$ distribution retrieved from the image. The quantitative description of each distribution is expressed in Appendix A. In order to obtain $P(x,y)$, the greyscale image matrix is normalized by dividing by the full intensity value of the image. $P(x)$ and $P(y)$ are obtained by summing along the y-axis and x-axis of the image, respectively, then normalizing by the total intensity of the image. $P(v)$ is obtained by taking a histogram of pixel values of an image.
\begin{figure}
  \centering
  \includegraphics[scale = 0.9]{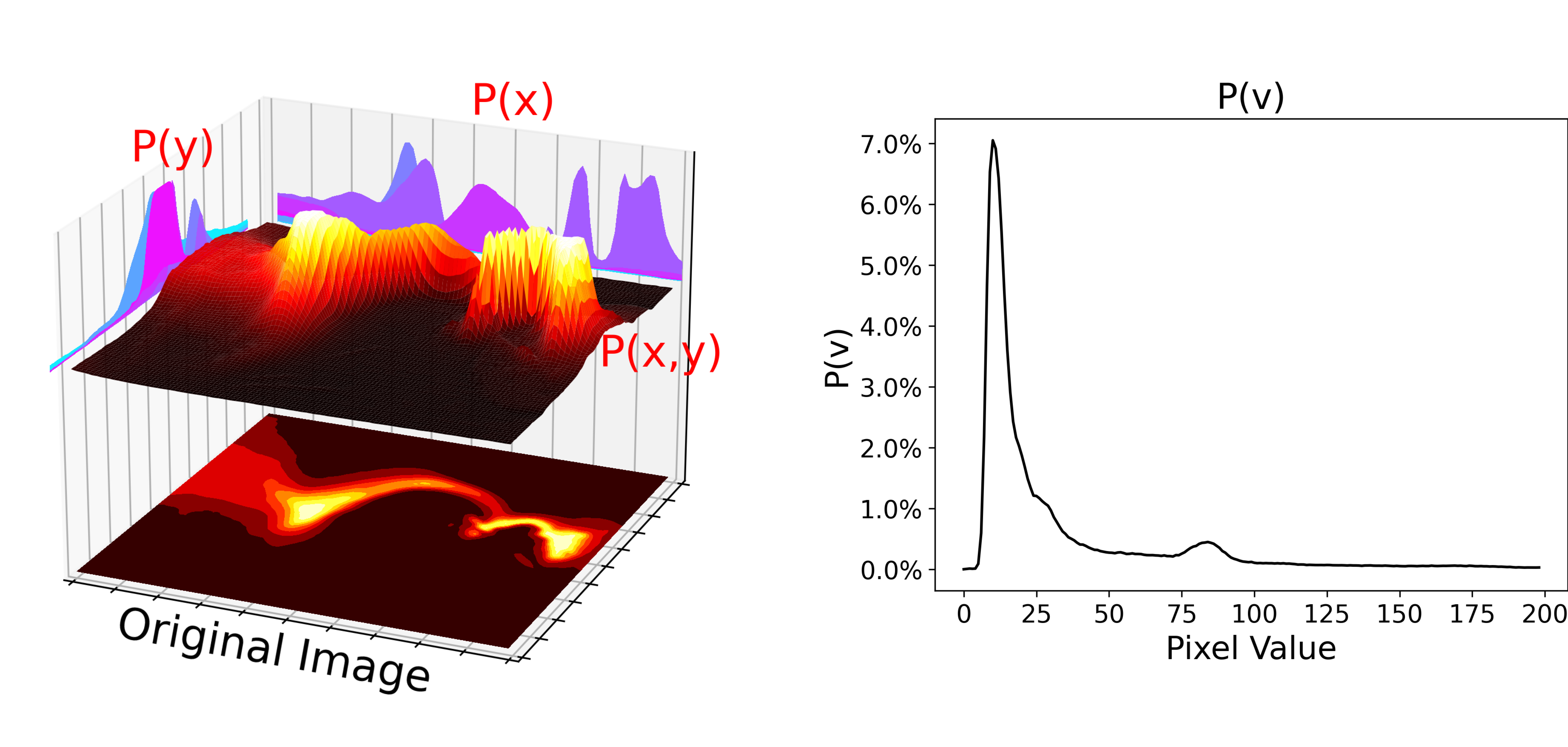}
  \caption{An example of an image from the kink class with the four distributions yielded to extract features. Note the image features a false colour map. $P(x,y)$, $P(x)$, and $P(y)$ are distributions of the intensity (left), while $P(v)$ is a histogram of pixels with particular values (right). Features are extracted from statistical and information theoretic quantities of these four distributions.}
\label{fig:3}
\end{figure}

With four distributions, we extract nineteen basic features from statistical and information theoretic quantities of the distributions, and we also include the total intensity value of each image as a feature. The quantitative description of each feature is expressed in Appendix A. The features are quite natural and easily calculable from an image and its distributions. They are as follows: the total intensity of the image (one feature), the raw mean and central variance of the $P(x)$, $P(y)$, and $P(v)$ distributions (six features), the third and fourth standardized moments of the $P(x)$, $P(y)$, and $P(v)$ distributions (six features), the entropy of each distribution (four features), the conditional entropies of the $P(x,y)$ distribution (two features), and the mutual information between the $P(x)$ and $P(y)$ distributions (one feature). These features can be computed on different pixel value ranges, allowing more features to be computed. We opt to compute the features on pixels valued from 200-300. 

In order to search for features that can distinguish classes, we create "cosine similarity matrices" which display the pair-wise average cosine similarity of classes. Generally, for each pair of two classes, we compute the cosine similarity ($c.s. = \hat{u}\cdot\hat{v}$) for each pair of vectors between the classes and calculate the average. We then display the results in a cosine similarity matrix, where the cross sections display the average cosine similarity value obtained between classes. It is expected that a proper feature set will have a matrix with diagonals, which include the similarity of each class with itself, with higher values compared to the off-diagonals, such that a matrix with diagonals of one and off-diagonals of negative one would be the perfect result. 

Specifically, we compute cosine similarity matrices yielded from all subsets of two features from the original twenty features and exhaustively search for a well behaved cosine similarity matrix indicating features geared towards a specific classification task. In Figure 4 we show a selection of a cosine similarity matrix from this search, with the mutual information of the $P(x)$ and $P(y)$ distributions, $I(x,y)$, and the mean of the $P(y)$ distribution for the features. From the cosine similarity of class one (spider) with itself compared to the cosine similarity of class one with the other four classes, these two features show potential for a binary, class one vs. non class one (spider vs non-spider) classification simply using comparison of feature vectors to a representative sample of class one vectors. Unfortunately, we do not see any matrices optimized for a full five-way classification of the data set from cosine similarity matrices of all possible feature vectors of length two, three, and four from our list of twenty features. Thus we conclude that these features are not good choices for our five-way plasma image classification goal. 
\begin{figure}
  \centering
  \includegraphics[scale = 0.5]{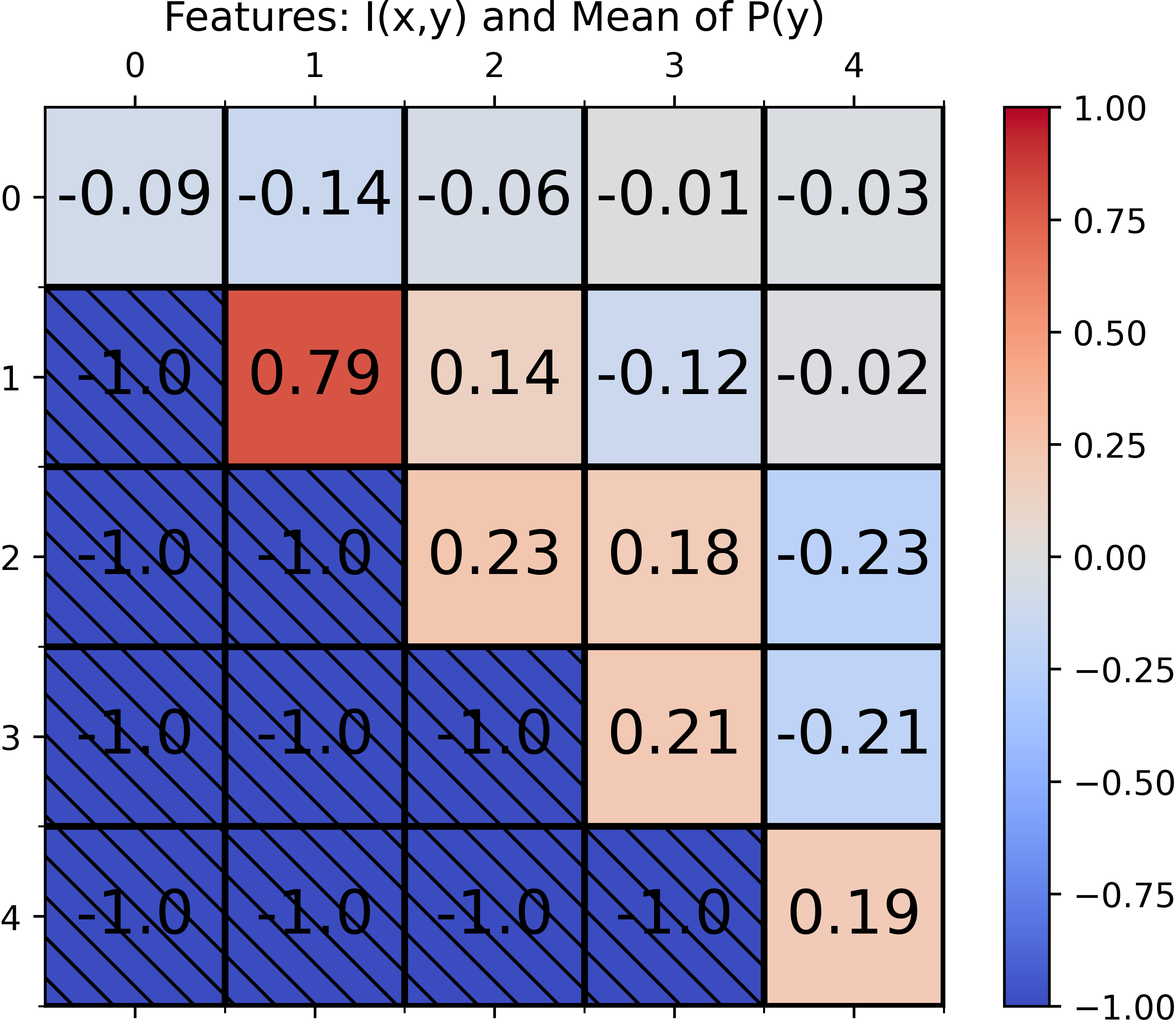}
  \caption{Example of a cosine similarity matrix indicating a potential feature selection for the task of class one or non class one (spider versus non-spider) binary classification. The two features used as components for vectors to construct the matrix were the mutual information, $I(x,y)$, and the mean of the $P(y)$ distribution. Classes are labeled as 0 through 4 corresponding to the classes shown in Figure 2. Since the matrix is normally symmetric, the bottom left half is set to the minimum value and crossed out.}
\label{fig:4}
\end{figure}
\subsection{A Binary Classification Algorithm Using Cosine Similarity of Feature Vectors}
We opt to construct a simple classifier using feature vector direction comparison for the following two reasons: one, to demonstrate the uses of cosine similarity for feature extraction by using the algorithm on the feature vectors that yielded Figure 4, and two, for the potential of the algorithm to be used with vectors that yield a more well behaved cosine similarity matrix and potential for multi-class classification.

Our binary classification algorithm is diagrammed in Figure 5 and outlined in the Algorithm 1 block. It is essentially a vector direction comparison algorithm. We first "train" the algorithm by selecting a representative sample of training images from the class. We extract feature vectors from each image in the training sample, and we compute the average cosine similarity of a large number of random pairs of feature vectors within the sample. Here the algorithm "learns" the similarity of images belonging to that class. The cosine similarity values calculated from "training" have a distribution with a training mean, $\mu$, and training standard deviation, $\sigma$. A test image, to be classified as in the class or not, is introduced to the algorithm, where its features are extracted and its cosine similarity is calculated with a satisfactory number (usually all) of the training samples. These new cosine similarity values have a test mean, $\mu^{*}$, which can be compared to the training mean. The training mean is representative of the average similarity an image in the class should have with its members, and the test mean represents the average similarity of the test image with the class. Classification is reduced to a threshold function, $\tau(\mu,\mu^{*},\sigma)$, which simply decides an image is in the class if $\mu^{*}$, the test mean, is greater than  $\mu - \gamma\sigma$. $\gamma$ is the only parameter featured in the algorithm.

\begin{algorithm}
\caption{: \textbf{Binary Classification With Cosine Similarity} \\
Given: iterations $N$, a number $\gamma$, a training set $U$, a test set $P$, and a vector-valued function $f(x)$}\label{alg:1}
\begin{algorithmic}
\State \:\:\:\:\:\:\:\:1.\:\:\:\:\textbf{for} $i$ in $N$: \Comment{"Training"}
\State \:\:\:\:\:\:\:\:2.\:\:\:\:\:\:\:\:\:\:\:\:$u \overset{\mathrm{iid}}{\sim}$ Unif$(U)$ ;  $v \overset{\mathrm{iid}}{\sim}$ Unif$(U)$
\State \:\:\:\:\:\:\:\:3.\:\:\:\:\:\:\:\:\:\:\:\:$x_i \xleftarrow{} \frac{f(u)\cdot f(v)}{\lVert f(u) \rVert \lVert f(v) \rVert}$
\State \:\:\:\:\:\:\:\:4.\:\:\:\:$\mu = \frac{1}{N}\sum\limits_{i=1}^N x_i$ ; $\sigma = \sqrt{\frac{\sum^{N}_{i=1} (x_i-\mu)^2}{N}}$ 
\State 
\State \:\:\:\:\:\:\:\:5.\:\:\:\:\textbf{for} $p$ in $P$: \Comment{"Testing"}
\State \:\:\:\:\:\:\:\:6.\:\:\:\:\:\:\:\:\:\:\:\:\textbf{for} $u$ in $U$:
\State \:\:\:\:\:\:\:\:7.\:\:\:\:\:\:\:\:\:\:\:\:\:\:\:\:\:\:\:\:$x_u \xleftarrow{} \frac{f(p)\cdot f(u)}{\lVert f(p) \rVert \lVert f(u) \rVert}$
\State \:\:\:\:\:\:\:\:8.\:\:\:\:\:\:\:\:\:\:\:\:$\mu^* = \frac{1}{\lVert U \rVert}\sum\limits_{u=1}^{\lVert U \rVert} x_u$ 
\State \:\:\:\:\:\:\:\:9.\:\:\:\:\:\:\:\:\:\:\:\:\textbf{if} $\mu^* > \mu - \gamma \sigma$:
\State \:\:\:\:\:\:\:\:10.\:\:\:\:\:\:\:\:\:\:\:\:\:\:\:\:\:\:\:$p$ is in class
\State \:\:\:\:\:\:\:\:11.\:\:\:\:\:\:\:\:\:\:\textbf{else}:
\State \:\:\:\:\:\:\:\:12.\:\:\:\:\:\:\:\:\:\:\:\:\:\:\:\:\:\:\:$p$ is NOT in class
\end{algorithmic}
\end{algorithm}

\begin{figure}
  \centering
  \includegraphics[scale = 0.25]{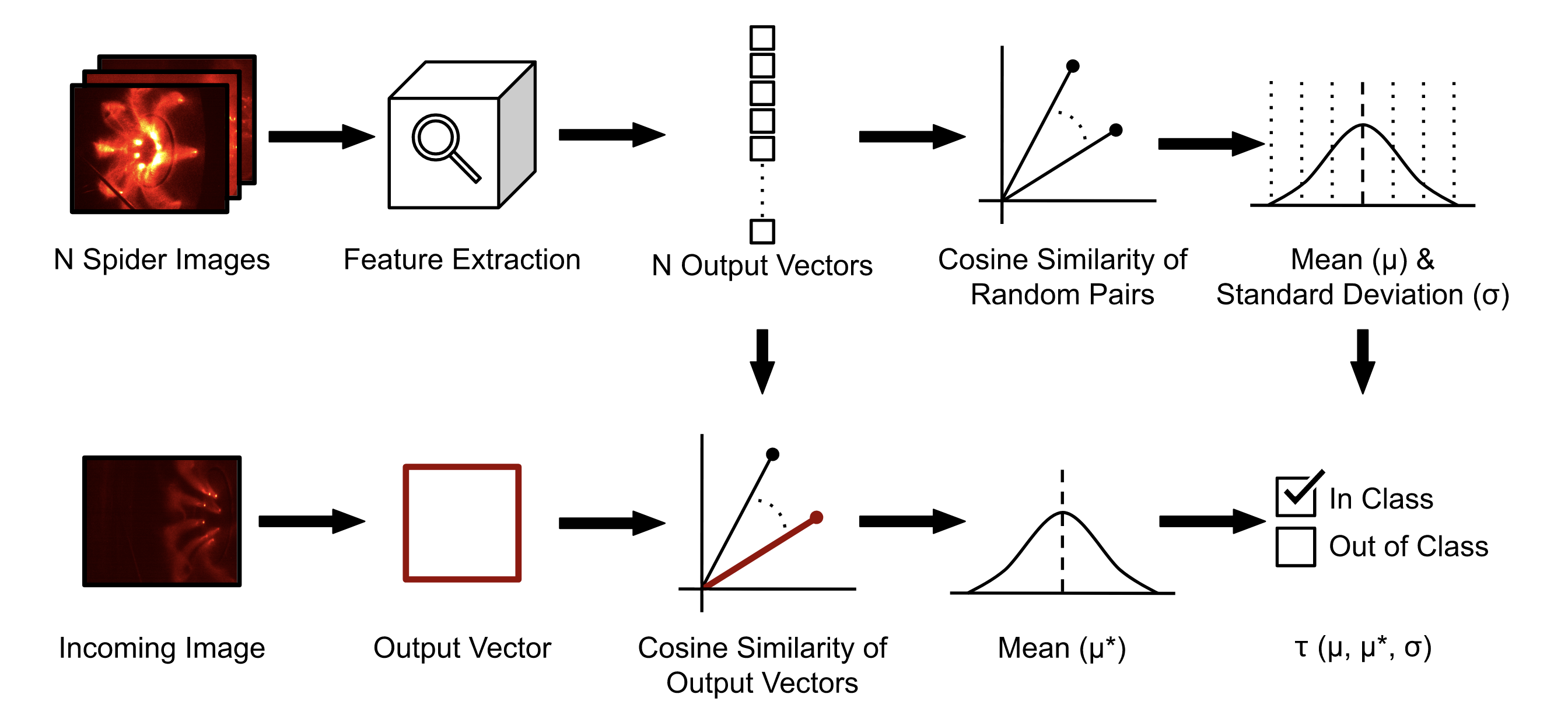}
  \caption{A diagram of the binary classification algorithm based on cosine similarity. We start with a "training" stage, such that enough vectors are sampled from the spider class an approximate average cosine similarity, $\mu$, of members in the class is retrieved, along with a standard deviation, $\sigma$, of the distribution of cosine similarity values in the class. With this information we introduce a test image to the algorithm and compute its average cosine similarity, $\mu^{*}$, with the training samples. Then a threshold function, $\tau (\mu, \mu^{*}, \sigma)$, which considers an image a member in the class if $\mu^*$ is greater than  $\mu - \gamma\sigma$, determines whether the image is in the spider class or not. Note the plasma images feature a false colour map. This algorithm is used to achieve notable results (shown in Figure 6) with the features from Figure 4.}
\label{fig:5}
\end{figure}

This binary classification method is used on the small data set with the same features from Figure 4 as a feature vector of length two for each image. We implement the method with fifty training samples from the spider class and fifty test samples from each class totalling 250 images to be classified. We choose to sample 1000 random pairs in the training set for the computation of $\mu$, and every vector in the training set is compared with the test images for the computation of $\mu^*$. $\gamma$ is chosen to be $0.01$ for optimal results. The results of spider versus non-spider binary classification using these presets are shown through an extended confusion matrix in Figure 6, where the algorithm performs surprisingly well despite its simplicity. Given a spider image, the algorithm classifies it as a spider with 80\% accuracy. For the remaining classes, the algorithm correctly classifies images as non-spider with 87.5\% average accuracy. In order to test the need for large amounts of training data, the same confusion matrix is calculated using 100 training pairs instead of 1000 with which the algorithm correctly classifies spiders and non-spiders with 94\% and 83\% accuracy, respectively. 
\begin{figure}
  \centering
  \includegraphics[scale = 0.9]{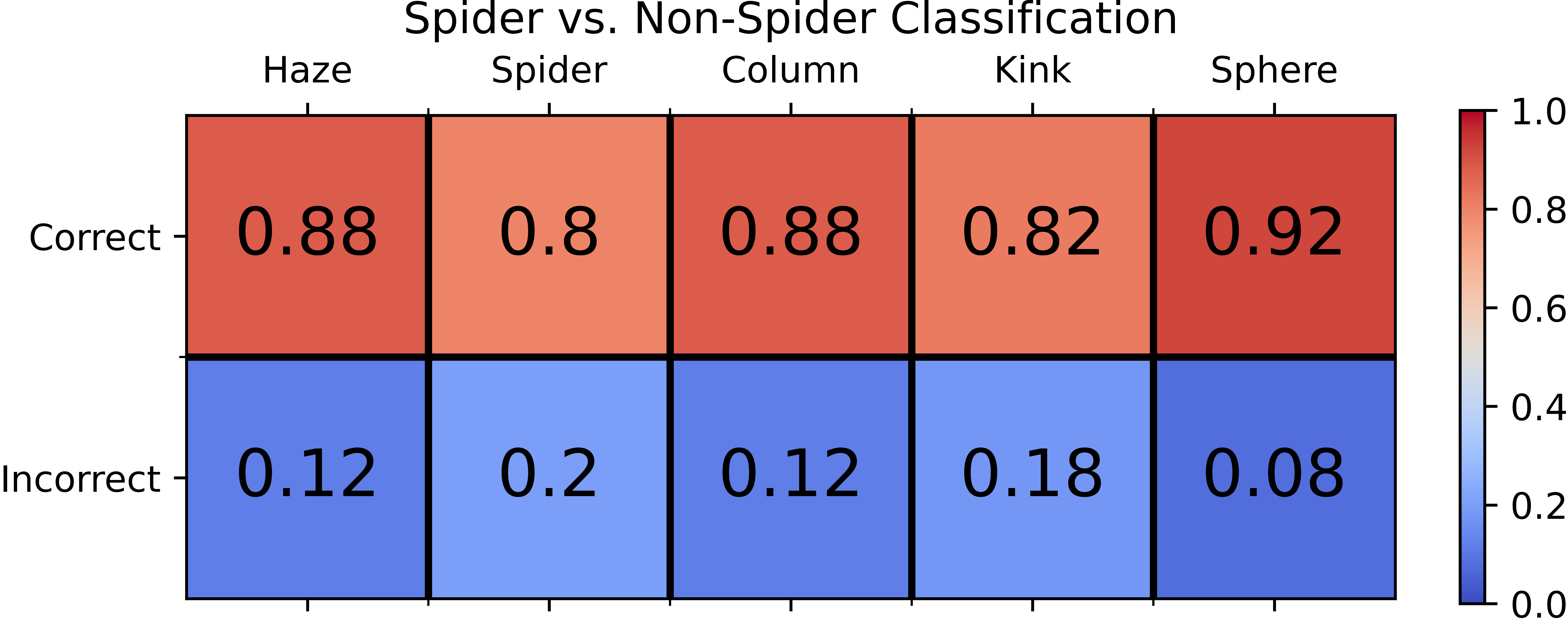}
  \caption{Confusion matrix of the results using the binary classification algorithm on the features from Figure 4 for spider versus non-spider classification. The matrix displays fractional accuracy for each class in the first row and the fractional inaccuracy for each class presented in the second row. The algorithm achieves an average accuracy of 86\%, and the total number of images classified is 250.}
\label{fig:6}
\end{figure}

\subsection{Transfer Learning Using Cosine Embedding and Cross Entropy Loss}
The results of the binary classification algorithm motivated us to conduct transfer learning on our plasma data set using both the cosine embedding loss function and the cross entropy loss function. This was done to attempt the full five-way classification of our data set and to probe how the training on our data set with different loss functions, specifically one related to cosine similarity, would allow for a well behaved cosine similarity matrix from output vectors of the network. 

We start AlexNet with the pretrained weights from pytorch \citep{pytorch}, which pertain to weights obtained upon the training of AlexNet on the Imagnet dataset \citep{imagenet}. The training, validation, and testing fractional split of our data is 0.5, 0.25, and 0.25. We use Amazon Web Service's Sagemaker for training along with hyperparameter optimization. We choose to optimize the learning rate (0.001 to 0.2), epoch number (5 to 100), and batch size (10 to 60) for cross entropy training and achieve five-way classification accuracies of 88\% for validation and 84\% for testing. For cosine similarity trained AlexNet, we choose the cosine embedding loss function, which works to maximize cosine similarity for objects in the same class and orthogonalizing objects in different classes. We implement the cosine embedding loss by using a hyperparameter, $\lambda$. In our classification, image labels are conventionally one-hot vectors, such as (0 0 1 0 0) corresponding to an image in class three from one to five. We then take predictions of the same form, where we simply take the maximum input in a vector, set it to one, and set the remaining inputs to zero. However, the cosine embedding loss between one-hot vectors results in gradients of one when the prediction is correct and zero when the prediction is incorrect. By setting all of the zeros in one-hot vectors to equal $\lambda$, we achieve the ability to conduct gradient descent, and thus, minimize the loss while also adding a parameter that tunes the direction of output vectors. If we set $\lambda$ to a constant negative one, we can perform gradient descent, but we heavily weight the direction of vectors through the inputs of negative one. We implement $\lambda$ with a range of  -0.4 to -0.05 while using the same hyperparameter ranges from cross entropy training for all other hyperparameters, and we achieve five-way classification accuracies of 89\% for validation and 82\% for testing. 

For a visualization of the activation differences between the trained versions of AlexNet, we employ the Gradient-Weighted Class Activation Mapping method \citep{gradcam}. The method highlights key areas of an image that allow the network to recognize a class by using the incoming gradients to the last convolutional layer of a network. Since the last convolutional layer contains the last instance of spatial information before being flattened in the fully connected layers, it ideally contains the most relevant class-specific feature activations which contribute to the fully connected layers and classification.
The method is implemented using readily available code\footnotemark[1]\footnotetext[1]{https://github.com/utkuozbulak/pytorch-cnn-visualizations}, with results shown in Figure 7. We note that the cross entropy trained model prioritized less of the total shapes for recognition of the column and kink images opposed to the cosine similarity trained model, which uses the majority of the kink and column shapes for their recognition. 
\begin{figure}
  \centering
  \includegraphics[scale = 1.5]{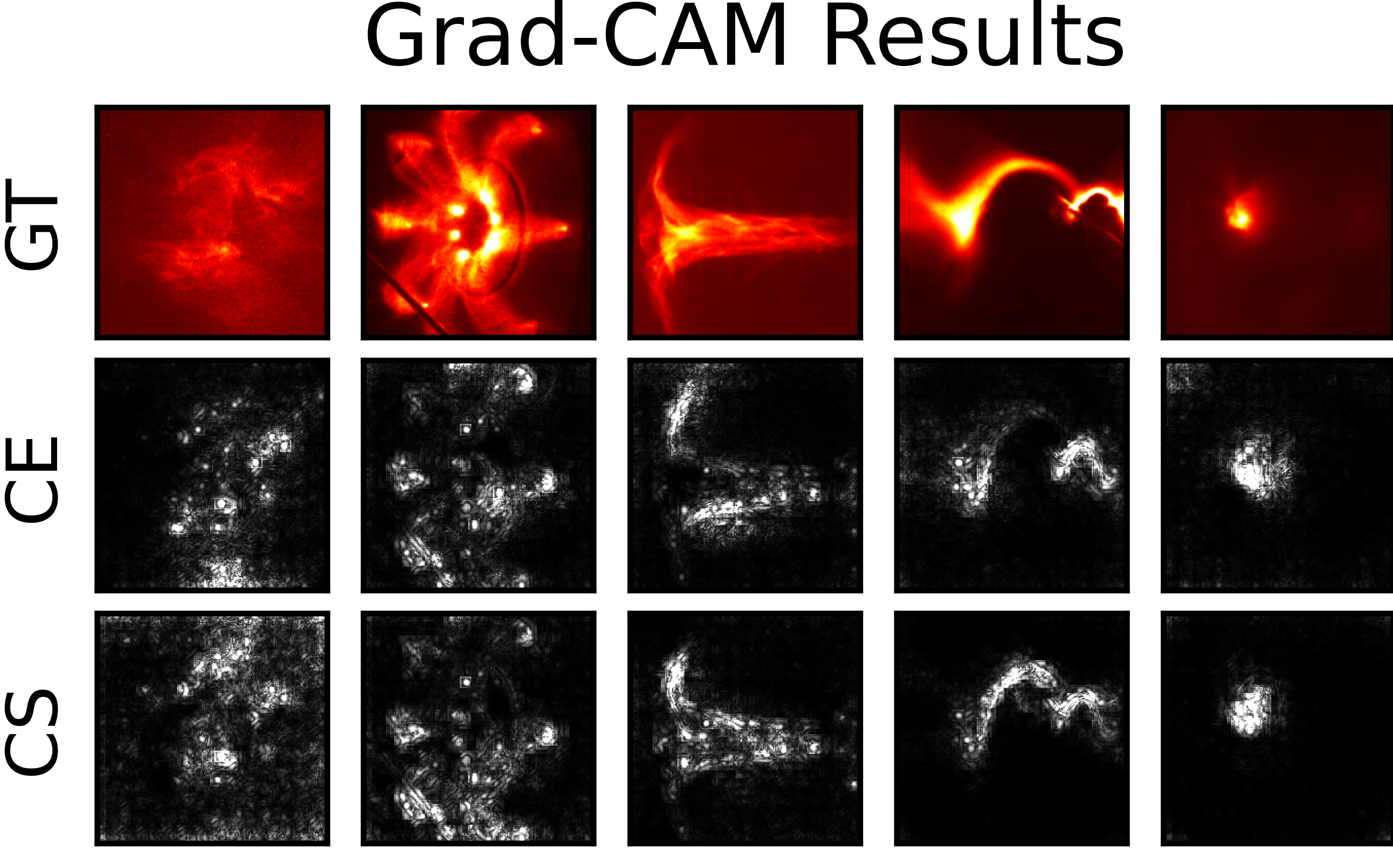}
  \caption{Differences in the activations between the cosine similarity trained AlexNet (labeled as CS) versus the cross entropy trained AlexNet (CE). The use of Gradient-weighted Class Activation Mapping (Grad-CAM) was performed on the five representative plasma images from Figure 2 for both versions of AlexNet. GT stands for "Ground Truth". The second and third rows display the outputs of Grad-CAM method on the first row for the corresponding models. Note the ground truth images are displayed using a false colour map.}
\label{fig:7}
\end{figure}

In addition to training using the cross entropy and cosine embedding losses, we also perform gradient descent using the cosine embedding loss on the final 4096 dimensional ReLU layer of AlexNet, with the aim of yielding a model that retrieves outputs with a well-behaved cosine similarity matrix of embedded vectors for use with our classification algorithm. The general training process is shown in Figure 8. 
\begin{figure}
  \centering
  \includegraphics[scale = 0.35]{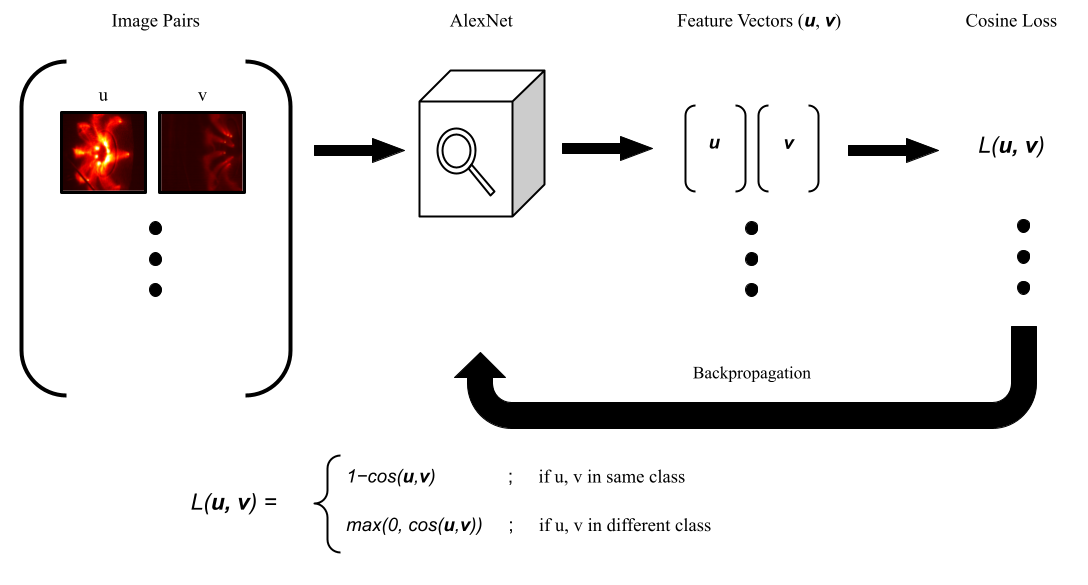}
  \caption{A diagram of the cosine embedding training on the final ReLU layer of AlexNet.}
\label{fig:8}
\end{figure}
We perform gradient descent with no classification results due to the final layer of the network no longer being a fully connected layer with outputs corresponding to classification. 

We then use the three trained models and compute cosine similarity matrices on output vectors from the final ReLU layer to display the differences in the cosine similarity of the final vectors before classification. Note that we perform this using the images from the test set only. Figure 9 shows these results, which demonstrate that conventional transfer learning of AlexNet does not necessarily include an optimization of cosine similarity between the final ReLU layer for images in the same class. Training by performing gradient descent on the embeddings themselves potentially gives the model the ability to perform with our vector direction comparison algorithm, so much so that vector direction comparison using cosine similarity could potentially be used to conduct full five-way classification of the test data set.
\begin{figure}
  \centering
  \includegraphics[scale = 0.20]{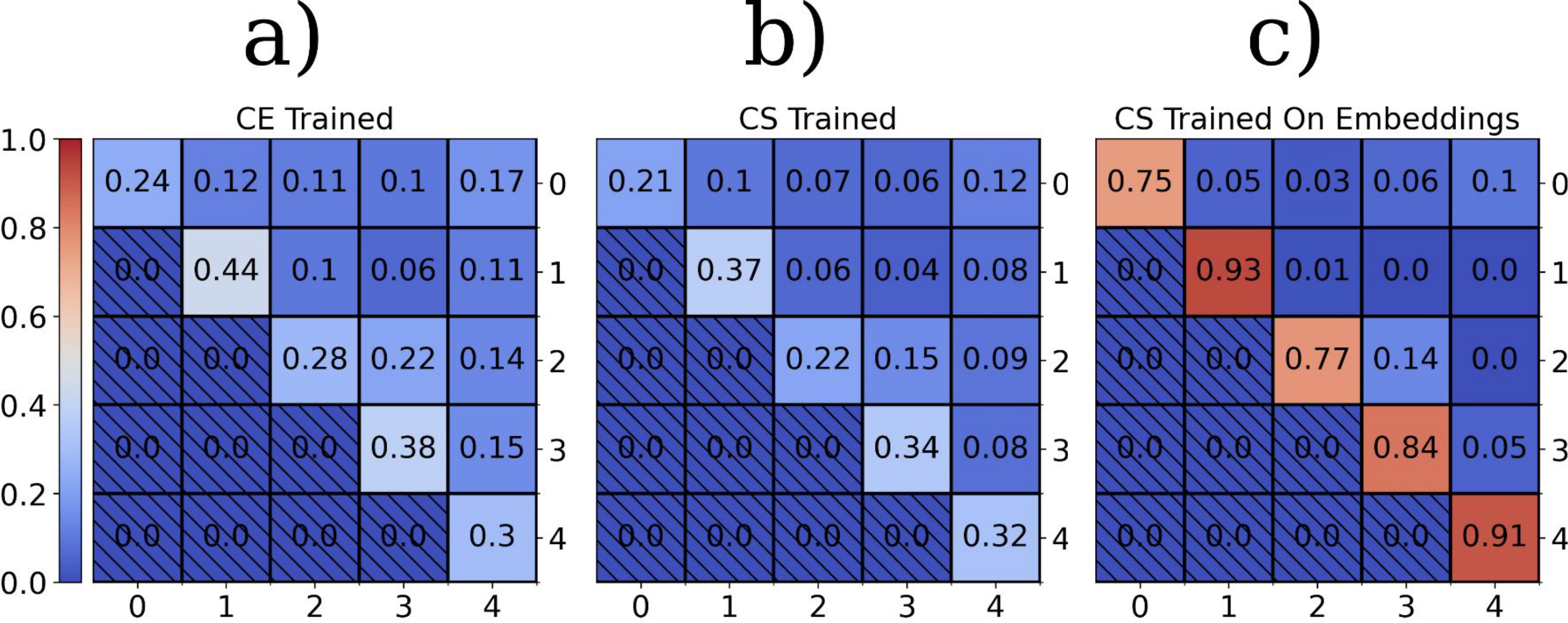}
  \caption{Cosine similarity matrices computed after performing transfer learning on AlexNet using a cosine embedding loss or cross entropy loss. Note the vectors used to construct the matrices were from the final ReLU layer of each network and that the matrices were computed using images from the test set only. The three sub-images show the cosine similarity matrices of the cross entropy trained (a), cosine embedding trained (b), and cosine embedding trained on the final ReLU layer (c) models. Since the matrices are normally symmetric, the bottom left half is set to the minimum value and crossed out.}
\label{fig:9}
\end{figure}

\subsection{Binary and Five-Way Classification Using Our Algorithm}
Since mere vector direction comparison seemed to work effectively with binary classification based on a cosine similarity matrix optimized towards one class (i.e. Figure 4), we expand the algorithm to include multi-class classification to use with the output vectors of AlexNet when trained on embeddings. Given Figure 9c, which shows the cosine similarity matrix obtained from the output vectors from this instance of AlexNet on the test set, the results should be notably better than the binary results for the engineered features (Figure 6). Now vectors in the same class, on average, are relatively aligned compared to vectors in different classes which have a comparably large angle between them. 

From our previous cosine similarity matrices obtained with engineered feature vectors and the traditionally trained cross entropy and cosine similarity models, there had yet to be a cosine similarity matrix indicating a strong distinction between kink and column images. Figure 9c displays a strong distinction of kink and column images. This is important since there is a key physical distinction between kinks and columns which depends on the ratio of the current in the plasma divided by the flux of the background poloidal magnetic field through the an inner plasma gun electrode \citep[see eq. 2.17]{bellan}. As an example of how the algorithm should perform better given a more well behaved set of vectors, we perform kink versus non-kink classification using the binary algorithm from section 2.3 with $\gamma$ set to $100$ for optimal results and 1000 randomly sampled training pairs. We display the results via confusion matrix in the left of Figure 10. Given a kink image, the algorithm notably classifies it as a kink with 92\% accuracy. For the remaining classes, the algorithm correctly classifies images in our test set as non-kink with 96\% average accuracy. It should also be noted that in distinguishing between stable columns and columns exhibiting kink instability, the algorithm is 93\% accurate. With ten training samples instead of 1000, the accuracies are robust at 92\% and 97\% for kink and non-kinks, respectively. 

Since a binary classification of vectors with well behaved cosine similarity achieved notable results, we modify the binary classification algorithm to perform multi-class classification based on an image being classified to the class it is most similar with. The modified algorithm is shown in the Algorithm 2 block. To change the algorithm for multi-class classification, we compute the average cosine similarity of one image with 50 samples from the training sets of each class to get five corresponding test means, $\mu_i$. We classify an image in the test set by assigning it to the class with the highest $\mu_i$. The results for full five-way classification of the test data (total of 285 images classified) are displayed in the right half of Figure 10 through a confusion matrix. The average accuracy for full, five-way classification of our test set is 92\%, roughly 10\% greater than our traditional transfer learning results of 82\% and 84\%. The results are identical when we compute the test means $\mu_i$ with 10 training samples from each class instead of 50.

\begin{algorithm}
\caption{: \textbf{Multi-Class Classification With Cosine Similarity} \\ Given: a number of classes $I$, a set of classes $C = \{C_i\}_{i \in I}$, a training set for each class $U_i$, a test set $P$, and a vector-valued function $f(x)$}\label{alg:2}
\begin{algorithmic}

\State \:\:\:\:\:\:\:\:1.\:\:\:\:\textbf{for} $p$ in $P$: 
\State \:\:\:\:\:\:\:\:2.\:\:\:\:\:\:\:\:\:\:\:\:\textbf{for} $i$ in $I$:
\State \:\:\:\:\:\:\:\:3.\:\:\:\:\:\:\:\:\:\:\:\:\:\:\:\:\:\:\:\:\textbf{for} $u$ in $U_i$:
\State
\State \:\:\:\:\:\:\:\:4.\:\:\:\:\:\:\:\:\:\:\:\:\:\:\:\:\:\:\:\:\:\:\:\:\:\:\:\:$x_u \xleftarrow{} \frac{f(p)\cdot f(u)}{\lVert f(p) \rVert \lVert f(u) \rVert}$
\State
\State \:\:\:\:\:\:\:\:5.\:\:\:\:\:\:\:\:\:\:\:\:\:\:\:\:\:\:\:\:$\mu_i = \frac{1}{\lVert U_i \rVert}\sum\limits_{u=1}^{\lVert U_i \rVert} x_u$ 
\State
\State \:\:\:\:\:\:\:\:6.\:\:\:\:\:\:\:\:\:\:\:\: $p \in C_i$; where $i = \underset{i}{\mathrm{argmax}}\:(\mu_i)$
\end{algorithmic}
\end{algorithm}

\begin{figure}
  \centering
  \includegraphics[scale = 1.0]{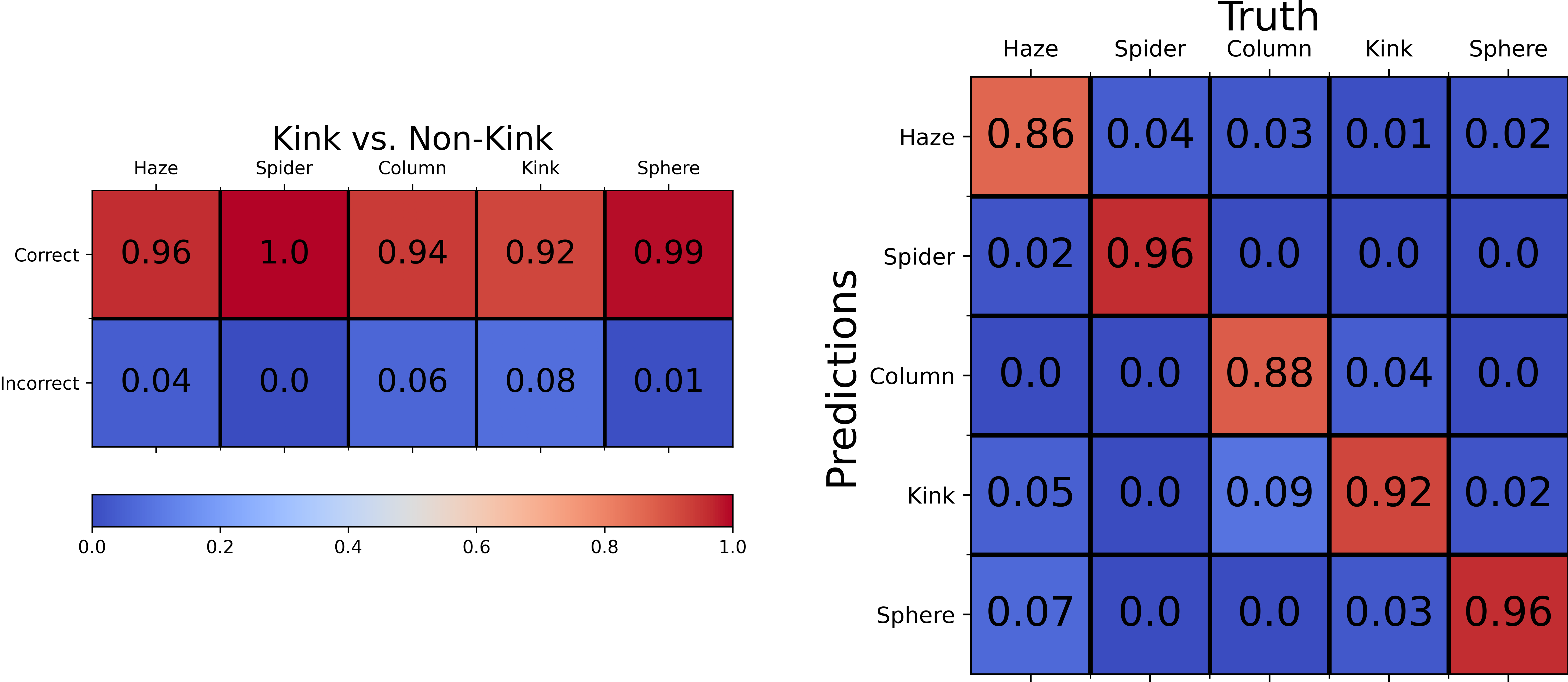}
  \caption{Confusion matrices of the test results using both the binary classification algorithm (left) and multi-class modification to the binary classification algorithm (right) on feature vectors obtained from AlexNet when trained on vector embeddings instead of the final layer. For the kink versus non-kink binary classification matrix the fractional accuracy for each class is displayed in the first row and the fractional inaccuracy for each class is presented in the second row. For the five-way classification results the diagonal values show the fractional accuracy of each class, while the off diagonal values show the fractional inaccuracies, displaying the fraction of each class in the x axis that was classified incorrectly as a class in the y-axis. The total amount of images classified for each case is 285.}
\label{fig:10}
\end{figure}

\subsection{Applying Previous Methods to the Large Data Set}

Our large data set from section 2.1 contains about 45000 unlabeled images. The instance of AlexNet trained using the cosine embedding loss on the final ReLU layer from section 2.4 is used to obtain feature vectors for each image in the large data set, and we classify the data set using the multi-class algorithm from section 2.5. We note the split of the data as 8\% haze, 15\% spider, 10\% column,  24\% kink, and 43\% sphere after classification is performed. 


With feature vectors for each image, we represent a significant portion of the data in two dimensions using T-distributed Stochastic Neighbor Embedding (tSNE) \citep{tsne}, which is shown in Figure 11. We note that the classes are joined in a common center point which could be representative of the fact each class is almost orthogonal to each other class in the feature space, and by using the vectors alone we do not get a representation based on the actual purpose of the network. Thus, we also perform tSNE on the normalized feature vectors in order to give a better idea of the grouping based on direction alone, which is shown in Figure 11.
\begin{figure}
  \centering
  \includegraphics[scale = 1.0]{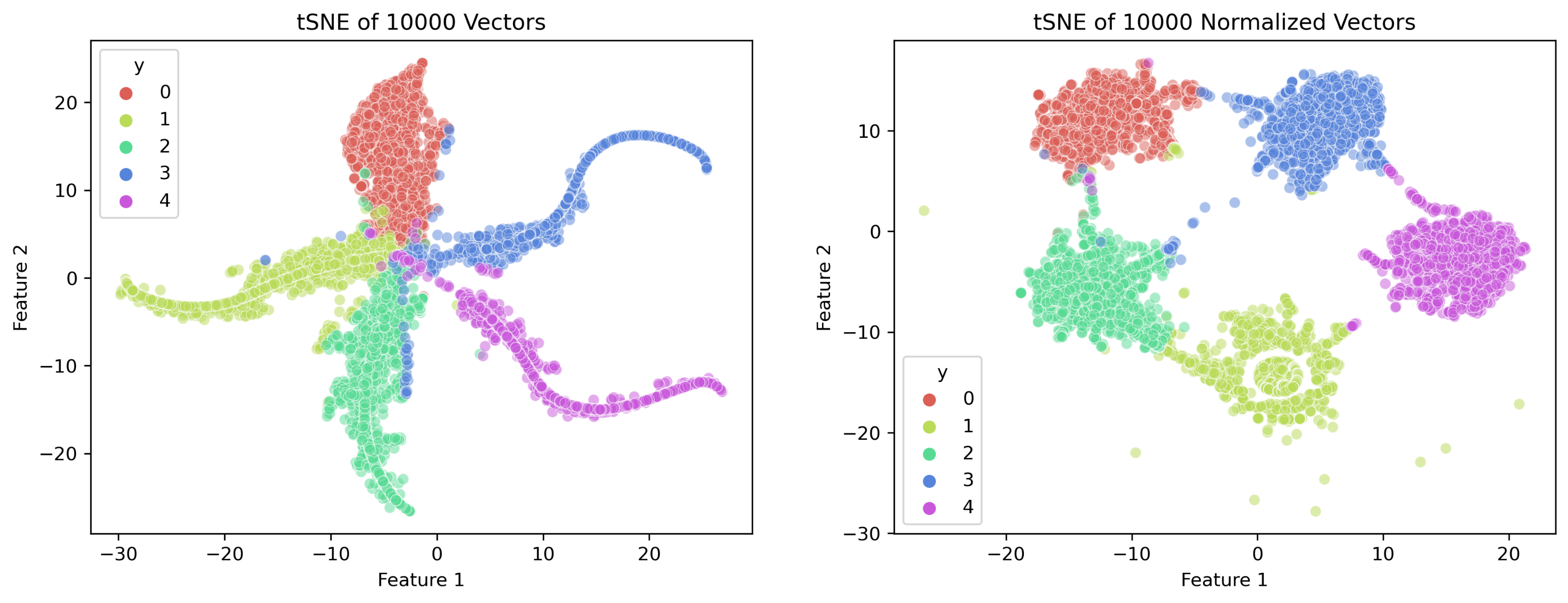}
  \caption{T-distributed Stochastic Neighbor Embedding (tSNE) for 2000 sampled images from each class (left). The tSNE parameters of perplexity and verobosity are set to 350 and 2, respectively. We note that when the 4096 dimensional vectors are reduced to two dimensions the data appears to be generally separated appropriately. We conduct the same tSNE with normalized feature vectors, where instead of five lines of classes, we see five clusters of classes. The classes are labeled as 0, 1, 2, 3, 4 corresponding to the labeling convention from the previous sections in this work.}
\label{fig:11}
\end{figure}

In order to investigate the distribution of data from each class, we plot the distribution of vectors about the average vector for a given number of samples in each class. An example of this can be seen in left half of Figure 12, for the kink class. We note that the distributions all heavily skew towards an angle of zero degrees, meaning the vectors for each class mostly populate a small angular region, with few outliers relatively misaligned with the class. In order to get an idea for the quality of accuracy for a class, despite not having ground truth, we display the closest images to the average vector, randomly sampled images from zero degrees to ten degrees, and randomly sampled images from forty five to fifty five degrees in the right half of Figure 12 for the kink class. Given the network was trained to align vectors in the same class, these qualitative results suggest the images with vectors further away from the average vector should become more frequently classified incorrectly or be images that fall outside of the five predefined class. 
\begin{figure}
  \centering
  \includegraphics[scale = 1.0]{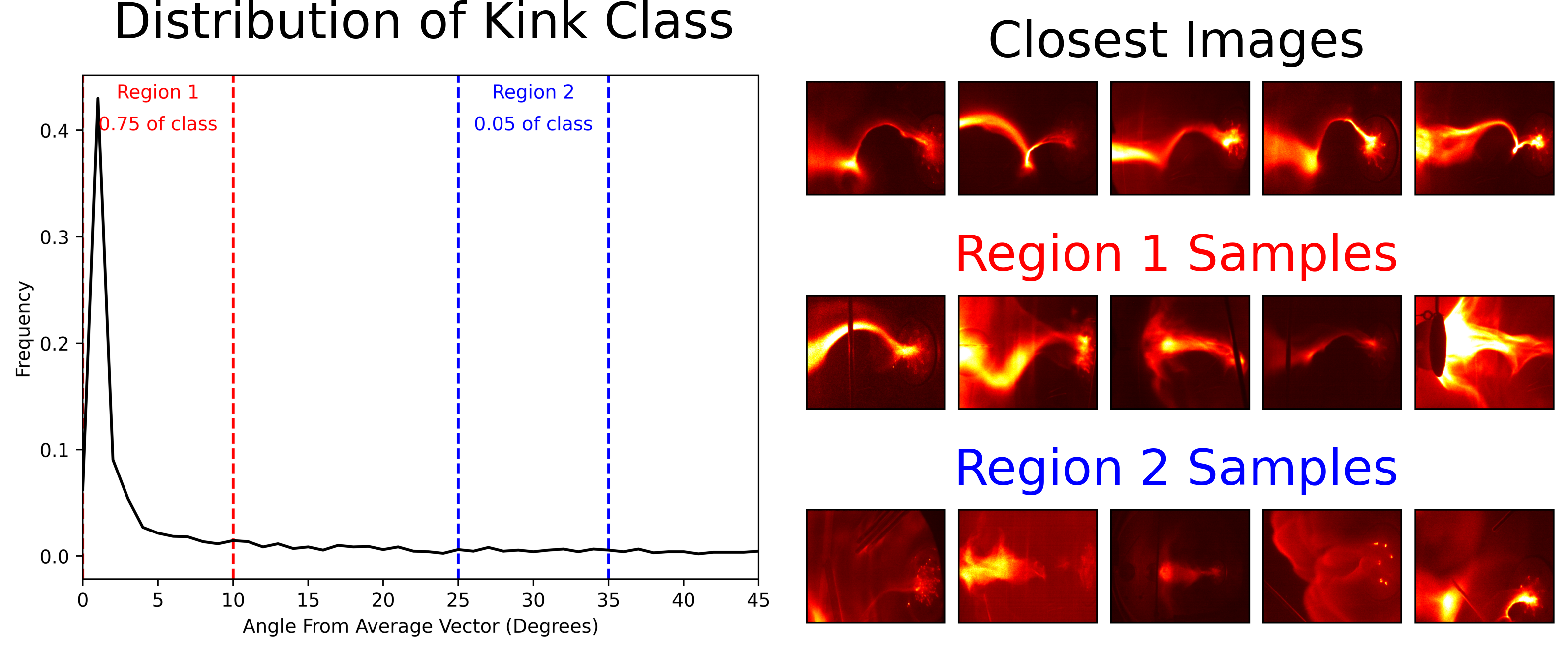}
  \caption{Distribution of 2000 randomly sampled vectors in the kink class about the average vector of the samples (left). The closest five images to the average vector, images randomly sampled from zero to ten degrees, and images randomly sampled from twenty five to thirty five degrees are displayed (right). 75\% of images in the class are within the zero to ten degree range, while the remainder of the data lies outside of this range.}
\label{fig:12}
\end{figure}


\section{Discussion}
This study is motivated by the goal of classifying a data set of plasma images from the Caltech Spheromak Experiment through feature extraction, transfer learning, and a vector direction comparison algorithm. We demonstrate the use of cosine similarity to select relevant features for binary classification with a well behaved cosine similarity matrix, which displays a representative average cosine similarity of vectors between each pair of classes. While the engineered features did not yield cosine similarity matrices indicating distinction among all five classes in our data set, simple vector direction comparison with no gradient descent based training was able to perform sufficiently well in the spider versus non-spider binary classification case. The idea of an extremely well behaved cosine similarity matrix inspired the use of AlexNet to train and yield one itself. We then use the embedding-trained model to conduct binary classification more accurately than the use of engineered features by 10\% from 86\% to 96\% simply because the feature vectors from this version of AlexNet were more well behaved. Based on the cosine similarity matrix in Figure 9c, when AlexNet is trained by the using the cosine embedding loss on the ReLU layer embeddings rather than the output layer, it seemingly groups image vectors in the same class together in the output space, so much so that representative average cosine similarity between test images in the same class is on average 0.84, corresponding to a relatively thin grouping for each class. Meanwhile the classes are separated enough for no class to share a similarity greater than 0.15 with another class. Thus, the image vectors from the final ReLU layer are divided into classes where each class is greatly separated by an angle relatively large compared to the angle between vectors in the class. This was exploited to expand our binary classification algorithm to a multi-class algorithm, and the five-way classification results are better than our traditional transfer learning results by about 10 percent from 82\% and 84\% to 92\%. When applying this uniquely trained version of AlexNet to the unlabeled data set, we can still see a propensity of the network to organize classes in thin groupings. In the ideal case with ground truth labels for a larger data set, the certainty an object belongs in the class as a function of angle away from the average vector can be probed quantitatively. This can potentially be used in the future to probe for sub-classes and outlier groupings corresponding to dense areas outside of the peak of a distribution. Preliminary developments of this idea are summarized in Fig. 11. 

While the method has been tested by our data sets, a possible new direction could be changing activation functions at the end of AlexNet to allow the outputs to access a larger domain, the use with more CNNs and different truncations of them, and the extension of the method for testing with more than five classes (and larger data sets in general). In our case using AlexNet and the final ReLU layer was sufficient for five classes.

\section{Summary}
The work performed in this study focuses on the classification of images from the Caltech Spheromak Experiment using a feature vector direction comparison algorithm and a modification of traditional transfer learning with AlexNet. With two engineered features, we demonstrate the vector direction comparison algorithm to perform binary classification with 86\% accuracy. We improve the accuracy of the algorithm to 96\% by using vectors from an instance of Alexnet trained to align vector embeddings in the same class while separating vector embeddings among different classes. Full five-way classification of the data set is performed with the output vectors from the embedding-trained AlexNet and a modification to the binary algorithm, which achieves a test accuracy of 92\%. This was larger than our test accuracies with transfer learning using the cross entropy loss (84\%) and the cosine embedding loss (82\%) on the final layer of the network for traditional five-way classification.
\section*{Funding}
M. F. was supported in part by a U.S. DoE Science Undergraduate Laboratory Internships (SULI) award. This work was also supported in part by the U.S. Department of Energy through the Los Alamos National Laboratory (ICF program). Los Alamos National Laboratory is operated by Triad National Security, LLC, for the National Nuclear Security Administration of U.S. Department of Energy (Contract No. 89233218CNA000001). 
\section*{Conflicts of Interest}
The authors declare no conflict of interest.
\appendix
\section{}\label{appA}
Section 2.2 uses engineered features, which are defined in the table below. 
\begin{table}
    \centering
    \renewcommand{\arraystretch}{2.0} 
     \begin{tabular}{p{5cm}||p{7cm}}
     \multicolumn{2}{c}{Given $G(x,y)$, the pixel value at $x, y$ and $g(v)$, the number of pixels with value $v$} \\
     \hline
     \textbf{Feature or Distribution} & \textbf{Equation}  \\
     \hline
     Intensity, $V$   &  $V = \sum\limits_{x = 1}^{}{\sum\limits_{y = 1}^{}{G(x,y)}}$  \\
     $P(x,y)$ Distribution &   $P(x,y) = \frac{1}{V}G(x,y)$  \\
     $P(x)$ Distribution &   $P(x) = \frac{1}{V}\sum\limits_{y = 1}^{}G(x,y)$  \\
     $P(y)$ Distribution &   $P(y) = \frac{1}{V}\sum\limits_{x = 1}^{}G(x,y)$  \\
     $P(v)$ Distribution & $P(v) = \frac{g(v)}{\sum\limits_{v = 1}^{}g(v)}$ \\
     Raw Mean, $\mu_1$ & $\mu_1 = \sum\limits_{x = 1}^{}x P(x)$ \\
     Central Variance, $\mu_2$ & $\mu_2 = \sum\limits_{x = 1}^{}(x-\mu_1)^2 P(x)$ \\
     nth Standardized Moment, $\mu_n$ & $\mu_n = \frac{1}{\sqrt{\mu_2}}\sum\limits_{x = 1}^{}(x-\mu_1)^n P(x)$ \\
     Entropy, $H(x,y)$ & $H(x,y) = \sum\limits_{x = 1}^{}\sum\limits_{y = 1}^{} -P(x,y) log_2(P(x,y))$ \\
     Conditional Entropy, $H(x|y)$ & $H(x|y) = H(x,y) - H(y)$\\
     Mutual Information, $I(x,y)$  & $I(x,y) = H(x) + H(y) - H(x,y)$ \\
     \hline
    \end{tabular}
    \caption{Equations to obtain features from section 2.2}
    \label{tab:my_label}
\end{table}
\newpage
\bibliographystyle{jpp.bst}
\bibliography{cs}
\end{document}